\documentclass[12pt]{article}
\usepackage{graphicx}
\usepackage{natbib}
\usepackage{url} 
\usepackage[latin1]{inputenc}
\usepackage{graphicx}
\usepackage{color}
\usepackage{geometry}  
\usepackage{parskip}   
\usepackage{amsmath, amssymb} 
\usepackage{setspace} 
\usepackage{booktabs} 
\DeclareMathOperator{\E}{\mathbb{E}}
\usepackage{float}
\usepackage{tabularx}
\usepackage{subcaption}
\usepackage{listings}
\usepackage{fancyvrb}
\usepackage[misc]{ifsym}

\newcommand{\bs}[1]{\ensuremath{\boldsymbol{#1}}}

\newcolumntype{Y}{>{\centering\arraybackslash}X}

\newcommand{\blind}{0}

\begin{document}

\def\spacingset#1{\renewcommand{\baselinestretch}%
{#1}\small\normalsize} \spacingset{1}

\if0\blind
{
  \title{\bf Detection of nonlinearity, discontinuity and interactions in generalized regression models}
  \author{Nikolai Spuck\thanks{
    The authors gratefully acknowledge the support by the German Research Foundation (DFG).}\hspace{.2cm},
    Matthias Schmid, and Moritz Berger \\
   \ \\
   Institute of Medical Biometry, Informatics and Epidemiology,\\
 Medical Faculty, University of Bonn
}
  \maketitle
} \fi

\if1\blind
{
  \bigskip
  \bigskip
  \bigskip
  \begin{center}
    {\LARGE\bf Title}
\end{center}
  \medskip
} \fi

\bigskip
\begin{abstract}
\noindent In generalized regression models the effect of continuous covariates is commonly assumed to be linear. This assumption, however, may be too restrictive in applications and may lead to biased effect estimates and decreased predictive ability. While a multitude of alternatives for the flexible modeling of continuous covariates have been proposed, methods that provide guidance for choosing a suitable functional form are still limited. To address this issue, we propose a detection algorithm that evaluates several approaches for modeling continuous covariates and guides practitioners to choose the most appropriate alternative. The algorithm utilizes a unified framework for tree-structured modeling which makes the results easily interpretable. We assessed the performance of the algorithm by conducting a simulation study. To illustrate the proposed algorithm, we analyzed data of patients suffering from chronic kidney disease.
\end{abstract}

\noindent%
\vfill

\newpage
\spacingset{1.75} 

\section{Introduction}

Generalized regression modeling is one of the most popular tools to analyse the association between an outcome variable of interest and one or several covariates. It is based on the assumption that the outcome variable follows a distribution from the exponential family, which makes it applicable to a wide range of outcomes on different scales like metrically scaled, categorical and binary variables. In generalized linear models (GLMs, \citealp{Fahrmeir2013}) one applies a link function to relate the expected value of the outcome to a set of covariates using a \emph{linear combination} of the covariate values. This means that the effects of continuous covariates are fitted by simple linear terms determined by a single coefficient, each. Thus, their effects can be easily interpreted (independent of the values of other covariates). Although this linear modeling approach is often considered the default and rarely questioned in practice, assuming linearity may often be too restrictive and misspecifying the functional form of a continuous covariate may lead to biased effect estimates and a decreased predictive ability \citep{Andersen2009}. While the application of variable selection methods in regression analysis (particularly in higher-dimensional settings) like the least absolute shrinkage and selection operator (LASSO, \citealp{Tibshirani1996}) and gradient boosting \citep{friedman2001} has become increasingly common, the importance of choosing an appropriate functional form of an effect is frequently neglected \citep{Sauerbrei2020}. As outlined in the following, there exist a lot of established approaches for modeling of continuous covariates (among others, categorization, structural breaks, polynomial regression, GAMs and CART) that go beyond classical GLMs. However, because each method exhibits specific benefits and drawbacks, the choice of the most appropriate one remains highly challenging.  To follow up on this issue, we propose a detection algorithm that examines various (nonlinear) modeling alternatives and guides practitioners to choose an appropriate one.  

A widely used nonlinear modeling approach for continuous covariates is \textit{categorization}. It avoids the need to make strong assumptions on the functional form of the effect and leads to results that appear to be easy to interpret by practitioners. Categorization, however, raises the question about the number of split points and where to place them. Although splitting into only two categories leads to the largest loss of information, dichotomization is widely popular. For instance, as stated by \citet{Nelson2017} dichotomization is frequently used in medical applications in order to stratify patients according to risk, make determinations about the necessity of additional diagnostic testing, or to allocate physician resources according to the patient's need. In the absence of predefined split points derived from domain knowledge, quantiles of the empirical distribution are often used, particularly in epidemiological research \citep{Sauerbrei2020}. Yet, the quantile-based approach is likely to select suboptimal split points since no information on the relationship between the continuous covariate and the outcome is used to guide split selection. In practice, choosing a split point based on a data-driven algorithm that takes the outcome variable into account (e.g.\@ by optimizing a certain metric) is usually more meaningful. Several data-driven methods for the selection of optimal split points (e.g.\@ by minimizing entropy or Gini impurity, or by maximizing a test statistic) are available, see, for example, \citet{Miller1982}, \citet{Strobl2007}, and \citet{Hastie2009}. 

Another way to model nonlinear relationships are \textit{structural breaks}, which are broadly applied in time series analysis and econometrics \citep{Andreou2009, Safikhani2021}. A model with structural breaks basically assumes a linear association between the continuous covariate and the outcome but allows that the regression coefficients (i.e.\@ intercept and slope) vary across the covariate space. In order to verify the presence of a structural break, \citet{Chow1960} introduced a test that examines whether the linear relationship between outcome and covariate changes at a prespecified covariate value. As with categorization, the breakpoints are rarely given exogenously in practice but are unknown and have to be estimated from the data. \citet{Zeileis2003} implemented approaches for detecting breakpoints and testing for structural breaks in time series and linear regression analyses. More recently, a review on the performance of methods for determining structural breaks in regression modeling was given by \citet{Guler2019}. Models with structural breaks constitute a special case of varying-coefficient models \citep{Hastie1993}, which allow the effects of continuous covariates to vary depending on the values of the same or other covariates (the so-called effect modifiers).

To capture smooth possibly nonlinear effects of continuous covariates on the outcome, polynomial terms (e.g.\@ quadratic or cubic terms) of the covariate values can be incorporated  in the model formula. 
A more flexible alternative are \textit{generalized additive models} (GAMs, \citealp{Hastie1990}) that allow to include continuous covariates with a smooth effect of unspecified functional form. In GAMs, a common way to specify the smooth functions is to use splines, which are represented by a weighted sum of basis functions, for example, by B-spline basis functions \citep{deBoor1978}. {Very flexible fits can be obtained by choosing a relatively large number of B-spline basis functions and include a term that penalizes differences between adjacent coefficients to prevent the estimated function from becoming too rough (P-splines; \citealp{Eilers1996}).
While classical GAMs offer great flexibility, they may not capture the structure of the data very well, if unknown interactions between covariates are present. This is because in additive models relevant interactions have to be specified in the model formula before fitting. An alternative regression approach that addresses this issue is \textit{recursive partitioning} or \textit{tree-based modeling}. The most popular version are classification and regression trees (CART) as proposed by \citet{Breiman1984}. CART recursively partition the covariate space into a set of disjoint hyperrectangles and in each hyperrectangle a simple model (e.g.\@ a constant) is fitted. Unlike generalized regression models, CART are able to automatically detect interactions without the need to include them in a prespecified model formula before fitting. Analogously to categorization, different metrics may be applied in order to select the optimal splitting rule (i.e.\@ variable and split point). In principle, categorization can be viewed as special case of CART where only one covariate is considered. CART can be visualized as hierarchical trees which makes the approach easily accessible for practitioners and simulatable \citep{Murdoch2019}. Yet, CART are incapable of modeling main effects and inherently assume nonlinearity, which may lead to a decreased predictive ability if linear effects are present in the data.

In this article, we propose an algorithm for the \textbf{de}tection of \textbf{n}onlinearity, \textbf{d}isconti-nuity and \textbf{i}nteractions (DENDI algorithm). The two-step algorithm utilizes tree-based splits which makes the resulting effects easily interpretable. More specifically, it indicates whether (i) linear effects are sufficient (indicating the use of a simple GLM), (ii) varying linear effects should be included in the model formula, (iii) one or several covariates exhibit non-linear effects (calling for the use of a GAM), or (iv) interaction effects occur in the data (hinting that the use of a nonparametric method like CART \citep{Breiman2001} may be beneficial). As described in the following, DENDI is based on a group of nested generalized regression models that can be embedded into the framework of tree-structured varying coefficient (TSVC) models \citep{Berger2019}.

We apply DENDI to data from the German Chronic Kidney Disease (GCKD) study \citep{Eckhardt2012, Titze2014}. The GCKD study is a prospective observational cohort study that includes patients with chronic kidney disease (CKD) of various aetiologies who are under nephrological care. The study aims at identifying relevant risk factors for different types and the severity of CKD. The study data includes many continuous variables such as age, body mass index (BMI) and biomarkers measured in urine/blood samples. As identifying an appropriate functional form for these variables may not only improve the predictive performance of a final model but also lead to a better understanding of the association between these markers and the specific form of CKD, detecting possible nonlinear and interaction effects constitutes a meaningful initial step in the analysis.

The remainder of this article is structured as follows: In Section~2, a group of nested generalized regression models is introduced, where the algorithm builds upon. We also describe how these models can be embedded into the TSVC framework. Subsequently, the DENDI algorithm is given in Section~3.  A summary of related approaches for the selection of functional forms is given in Section~4. To asses the performance of the proposed algorithm, we performed a simulation study. The results are presented in Section~5. In Section~6, we illustrate the application of the DENDI algorithm by an analysis of the GCKD study data. Finally, advantages and limitations of our algorithm are discussed in Section~7.

\section{A group of generalized regression models}

We consider generalized regression models, where the expectation of an outcome $y_i,\;i=1,\hdots,n$ is linked to a vector of $p$ covariates $\bs{x}_i = (x_{i1},...,x_{ip})^\top$ in the form 
\begin{equation}
\label{generalized}
\E(y_i|\ \bs{x}_i) = g^{-1}(\eta (\bs{x}_i)), \quad i = 1,...,n\,,
\end{equation}
where $g(\cdot)$ denotes a suitable link function and $\eta(\cdot)$ denotes the predictor function. In the following we introduce five nested models in which the predictor $\eta(\cdot)$ takes different forms as shown in Figure~\ref{Fig1}. 

First, let us focus on one continuous covariate $x_j$. Assuming that the effect of~$x_j$ on the outcome is simply \textit{linear} yields the model with predictor function
\begin{equation}
\label{linear}
\eta(\bs{x}_i) = \beta_0 + \beta_j \, x_{ij}\, ,
\end{equation}
 where $\beta_0$ is the intercept and $\beta_j$ is the linear regression coefficient. In place of the linear term $\beta_j\, x_{ij}$, one can consider a predictor function with \textit{piecewise constant} effect, which has the form 
\begin{equation}
\label{pc}
\eta(\bs{x}_i) = \beta_0 +  \gamma_j \, I(x_{ij} > c_j)\, ,
\end{equation}
where $I(\cdot)$ denotes the indicator function, $c_j$ is a split point in $x_{j}$ and $\gamma_j$ is the corresponding regression coefficient. Note that the piecewise constant model in~\eqref{pc} involving one split point is equivalent to dichotomization with regard to~$c_j$ and may also be written as $ \gamma_{j1}\, I(x_{ij} \leq c_j) +  \gamma_{j2}\, I(x_{ij} > c_j)$. If both models ~\eqref{linear} and~\eqref{pc} are not fully appropriate to capture the predictor-response relationship, they can be extended by one additional term. A more complex model using an \textit{additive combination} of the linear and piecewise constant effect yields the predictor function 
\begin{equation}
\label{add}
\eta (\bs{x}_i) = \beta_{0} + \beta_{j} \, x_{ij} +  \gamma_{j} \, I(x_{ij}>c_j)\, .
\end{equation}
Note that both, the linear and  the piecewise constant model are nested in model~\eqref{add}. It represents a structural break with regard to the intercept, while the slope $\beta_{j}$ remains the same across the entire covariate space. When using a \textit{multiplicative combination} of the linear and piecewise constant effect, the predictor function is given by 
\begin{align}
\label{mult}
\eta(\bs{x}_i) &= \beta_0^*+ \beta_{j1} \,I(x_{ij} \leq c_j)\, x_{ij} +\beta^*_{j2}\, I(x_{ij}>c_j)\, x_{ij}\, \nonumber\\
&=\beta_0 + \beta_{j1}\, x_{ij}  + \beta_{j2}\, I(x_{ij}>c_j)\,( x_{ij} - c_j) \,,
\end{align}
where $\beta_{0}^*= \beta_0 -\beta_{j2}I(x_{ij}>c_j)c_j$ and $\beta_{j2}=\beta^*_{j2}-\beta_{j1}$. From the second equation in~\eqref{mult} it is seen that the linear model is nested in this model, as setting $\beta_{j2} = 0$ yields model~\eqref{linear}. Here, the predictor function represents a structural break with regard to the slope. Note that introducing the subtrahend $c_j$ ensures continuity of the fitted function (see also Figure~\ref{Fig1}). Finally, we consider an extension allowing for an \textit{additional split} in $x_j$, which has the form
\begin{equation} 
\label{tree}
\eta(\bs{x}_i) = \\
\begin{cases}
\beta_0 + \gamma_{jr}\, I(x_{ij} > c_{j}) \\~~~
+ \gamma_{j\ell}\,I(x_{ij} \leq c_j \land x_{ij}>c_{j\ell})\,,\quad \text{if split in}\;\{x_{ij} \leq c_j\}\,, \\
\beta_0 + \gamma_{j\ell}\, I(x_{ij} \leq c_{j})  \\~~~
+ \gamma_{jr}\, I(x_{ij} > c_j \land x_{ij}> c_{jr})\, ,\quad\text{if split in}\;\{x_{ij} > c_j\}\,,
\end{cases}
\end{equation}
where $c_{j\ell} < c_j$ and $c_{jr} > c_j$.
This model is based on~\eqref{pc} and takes two different forms depending on whether the left node or the right node is chosen for the second split. Figure~\ref{Fig1} shows an illustration of the five different models \eqref{linear} to \eqref{tree} and their nested structure.  

\begin{figure*}[!t]
\centering
\includegraphics[width = 11.9cm, trim = 0cm 0cm 0cm 0cm]{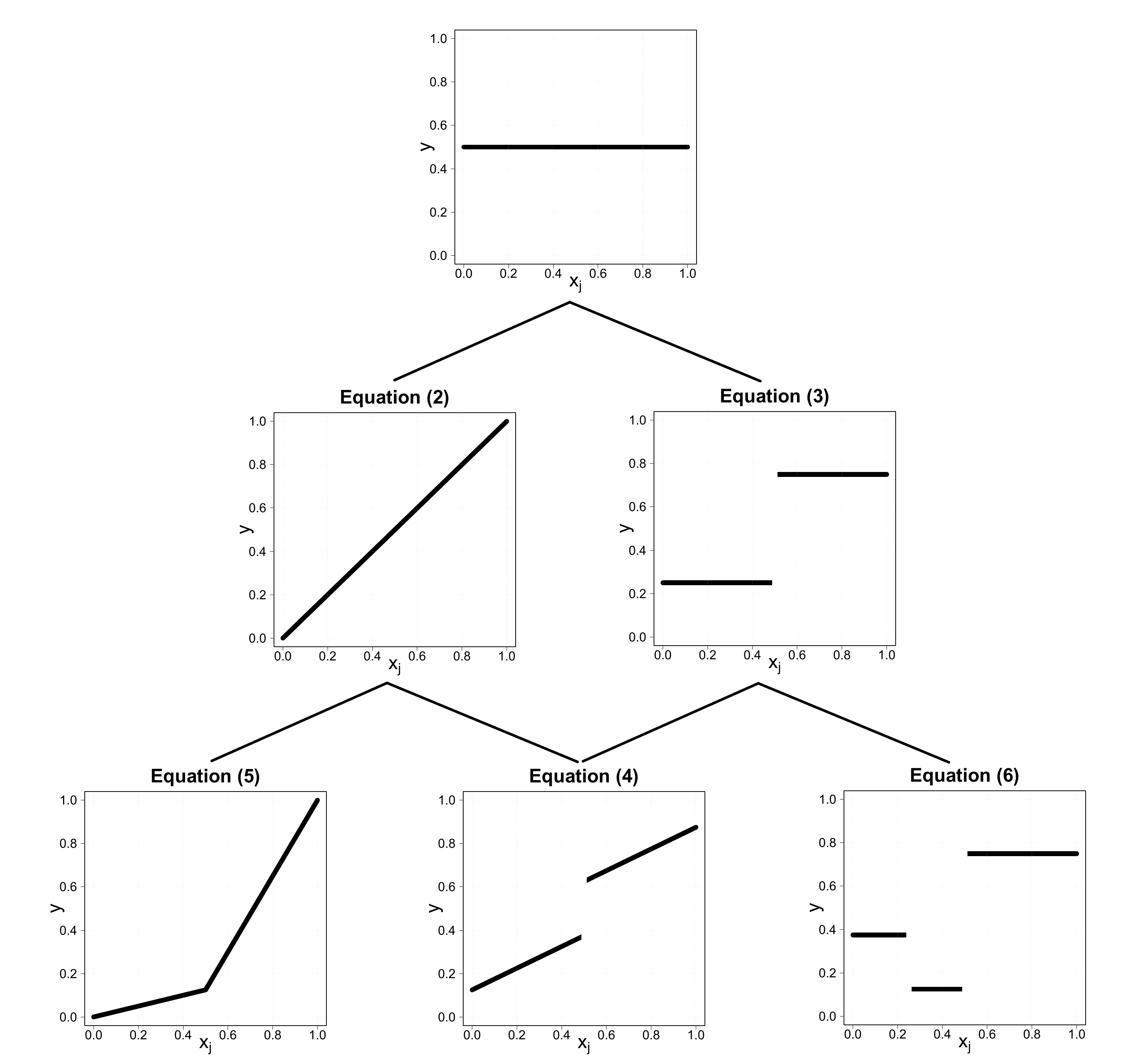}
\caption{Illustration of the group of five nested models. The figure shows a tree structure that describes the relationships between the models considering one covariate $x_j$. The models in the parent nodes are nested in each of the models in their child nodes.}
\label{Fig1}
\end{figure*}

\subsection*{Several Covariates}

In the presence of multiple continuous covariates $x_1,\hdots,x_p$ each corresponding part of the predictor function can take the form as given by \eqref{linear} to \eqref{tree}. Furthermore, the multiplicative effect in \eqref{mult} and the tree-structured effect in \eqref{tree} allow for an interaction between two covariates at a time. Let us consider two covariates $x_j$ and $x_k$, then a multiplicative combination of effects yields the predictor function 
\begin{equation}
\label{mult2}
\eta(\bs{x}_i) = \beta_0 + \beta_{j1}\, x_{ij}  + \beta_{j2}\,I(x_{ik}>c_k)\,x_{ij}\,.
\end{equation}
Importantly, the model in~\eqref{mult2} corresponds to a varying-coefficient model, where the linear effect of $x_j$ is modified by $x_k$. A tree-structured model with a first split in $x_j$ and a second split in $x_k$ (in the left node) has the form  
\begin{equation} 
\label{tree2}
\eta(\bs{x}_i) = \beta_0 + \gamma_{jr}\, I(x_{ij} > c_{j})+ \gamma_{j\ell}\, I(x_{ij} \leq c_j \land x_{ik}>c_{k})\,.
\end{equation}
The predictor function in~\eqref{tree2} equals a tree consisting of three leaves (where the lower left node serves as reference) and represents an interaction between $x_j$ and~$x_k$. 

\subsection*{Tree-Structured Varying Coefficients}

The detection algorithm (DENDI) introduced in the next section is based on the five univariable models~\eqref{linear} to~\eqref{tree} and the two bivariable models~\eqref{mult2} and~\eqref{tree2}. Technically, these models can all be embedded into the framework of TSVC models \citep{Berger2019}. To obtain coefficient estimates we make use of the eponymous R-add-on package \citep{berger2018TSVC}. In general, varying-coefficient models by \citet{Hastie1993} have the form
\begin{equation}
\label{vc}
\eta (\bs{x}_i) = \beta_{0} + \beta_{1}^V(v_{i1})\, x_{i1} +...+ \beta_p^V (v_{ip}) \, x_{ip}\, ,
\end{equation}
where $v_{i1},\hdots,v_{ip}$ denote (additional) covariates that serve as effect modifiers and change the linear effects of $x_{i1},\hdots,x_{ip}$ by an unspecified functional form $\beta_j^V(\cdot)$. The specification of models~\eqref{pc} to~\eqref{tree2} within the TSVC framework is given in Table~\ref{TSVCforms}.
\begin{table}[!t]
\caption{Tree-structured varying coefficient models. The table shows how the group of nested regression models is embedded into the framework of TSVC models.\vspace{0.2cm}}
\label{TSVCforms}
\begin{center}
\begin{footnotesize}
\begin{tabularx}{\textwidth}{X|ll}
\toprule 
\small
Model & \multicolumn{2}{l}{Specification within TSVC}\\
\hline
Predictor \eqref{pc}& $\eta(\bs{x}_i)=\beta_0^V(x_{ij})$ & $\beta_0^{V}(x_{ij}) =  \beta_0 + \gamma_1 I(x_{ij} > c_j)$\\
Predictor \eqref{add} & $\eta(\bs{x}_i)=\beta_0^V(x_{ij})+\beta_{j}\, x_{ij}$ & $\beta_0^V(x_{ij}) = \beta_{0} + \gamma_{j} I(x_{ij} > c_j )$\\
Predictor \eqref{mult} &$\eta(\bs{x}_i) = \beta_0^* + \beta_j^V(x_{ij})\, x_{ij}$ & $\beta_j^V(x_{ij}) = \beta_{j1}I(x_{ij}\leq c_j) + \beta_{j2}^* I(x_{ij}>c_j)$\\
Predictor \eqref{mult2} &$\eta(\bs{x}_i) = \beta_0 + \beta_j^V(x_{ik})\,x_{ij}$ & $\beta_j^V(x_{ik}) = \beta_{j1} + \beta_{j2} I(x_{ik}>c_k)$\\
Predictor \eqref{tree} & $\eta(\bs{x}_i)=\beta_0^V(x_{ij})$ & $\beta_0^V(x_{ij}) = \beta_0 + \gamma_{jr}\,I(x_{ij} > c_{j})$ \\ 
&&$\hphantom{\beta_0^V(x_{ij}) =\beta_0}+\gamma_{j\ell}\,I(x_{ij} \leq c_j \land x_{ik}>c_{jr})$\\
Predictor \eqref{tree2} & $\eta(\bs{x}_i)=\beta_0^V(x_{ij},x_{ik})$ & $\beta_0^V(x_{ij},x_{ik}) = \beta_0 + \gamma_{jr}\,I(x_{ij} > c_{j})$ \\
&&$\hphantom{\beta_0^V(x_{ij},x_{ik}) = \beta_0}+ \gamma_{j\ell}\,I(x_{ij} \leq c_j \land x_{ik}>c_{k})$\\
\bottomrule
\end{tabularx}
\end{footnotesize}
\end{center}
\end{table} 

\section{Algorithm}

Based on the group of models introduced in the previous section we propose the two-step DENDI algorithm that examines these modeling alternatives and chooses the one that maximizes predictive performance. More specifically, we compute the predicted log-likelihood of the models using leave-one-out cross validation (LOOCV). In addition, we apply the so-called ``one standard error rule'' (1SE rule), which is an established strategy for the selection of tuning parameters in regularized regression \citep{Chen2021}.

Let us again focus on one continuous covariate $x_j$, only. In the first step of the algorithm, the two models on the second level of the tree structure in Figure~\ref{Fig1}, namely the linear model \eqref{linear} and the piecewise constant model \eqref{pc} are evaluated. Among these two models, the model with the largest predictive log-likelihood (averaged over all observations) is selected and compared to the null model (with intercept $\beta_0$ only). For this, the 1SE rule is applied as follows: Let $p\ell_{1}^{[0]},\hdots,p\ell_{n}^{[0]}$ be the predicted log-likelihood values of the null model obtained from LOOCV, then the average predicted log-likelihood of the selected model~$p\ell^{[1]}$ is compared~to  
\begin{align*}
\frac{1}{n}{\sum_{i=1}^{n}}{\,p\ell_{i}^{[0]}}+\sqrt{\frac{\text{Var}(p\ell_{1}^{[0]},\hdots,p\ell_{n}^{[0]})}{n}}=p\ell^{[0]}+SE^{[0]}\,,
\end{align*}
that is, to the average predictive log-likelihood of the null model and its standard error. If $p\ell^{[1]} > p\ell^{[0]}+SE^{[0]}$, the selected model is confirmed and the algorithm continues with step 2. Otherwise, no effect of $x_j$ was found and algorithm is terminated with the final output being the null model. 

In the second step of the algorithm, the models on the third level of the tree structure in Figure~\ref{Fig1} are considered. If a linear effect was selected in step 1, the models with an additive combination of effects~\eqref{add} and a multiplicative combination of effects~\eqref{mult} are evaluated. Otherwise, if a piecewise constant effect was selected in step~1, the model with an additive combination of effects~\eqref{add} and the tree-structured model~\eqref{tree} are evaluated (as illustrated in Figure~\ref{Fig1}). In the same way as in step~1, the algorithm firstly computes the average predictive log-likelihood values of the two models using LOOCV, and secondly compares the better performing model to the simpler (linear or piecewise constant) model applying the 1SE rule.   

\sloppypar{DENDI also allows for an additional vector of confounding variables $\bs{z}_i = (z_{i1},...,z_{iq})^\top$ (e.g.\@ dummy-coded binary or categorical covariates), one may want to adjust for when selecting an appropriate functional form for $x_j$. To do so, in each fitting step of the algorithm the predictor of the model is complemented by the linear term $\bs{\delta}^\top \bs{z_i}$, with regression coefficients $\bs{\delta} = (\delta_1,..., \delta_q)^\top$. }

If not only one but multiple continuous covariates $x_1,\hdots,x_p$ are of interest, the DENDI algorithm additionally allows to investigate whether interactions are present in the data, and needs to be slightly adapted. In a multivariable scenario, step~1 of the algorithm is first performed for all $p$ covariates separately. All effects that were detected according to LOOCV and the 1SE rule (linear or piecewise constant) are then included in step~2. That is, when investigating the modeling alternatives for one covariate $x_j$ on the third level of the tree structure in Figure~\ref{Fig1}, each model is adjusted for all effects of the other covariates selected in step 1. Otherwise, interactions might be falsely detected, just because relevant main effects are neglected.

Following the TSVC approach by \citet{Berger2019}, when fitting models \eqref{pc} to \eqref{tree2} the split points are selected by the deviance (that is, minus two times the log-likelihood). More specifically, for one covariate $x_j$ a fixed number of splits points (defined by quantiles of $x_j$) are examined and the split point that yields the smallest deviance is used for splitting. Note that, when the piecewise constant effect was selected in step~1 of the algorithm, the corresponding split point is kept when examining the modeling alternatives in step 2. When fitting models~\eqref{tree} and \eqref{tree2}, the best split is selected among all possible split points and among the two currently built nodes. 

A detailed description of the DENDI algorithm given multiple continuous covariates and categorical confounders is given in the Supplementary Material.

\section{Related approaches}

Alternative approaches for choosing suitable functional forms of covariate effects in regression analysis have been proposed by a number of researchers. In the following, we give an overview of these approaches and discuss their advantages and limitations compared to DENDI.

An approach for the detection of nonlinearity, which is also based on trees, was proposed by \citet{Su2008}. They consider scenarios, where an outcome variable~$y$ is related to a mixture of continuous and categorical covariates $x_1,...,x_p$, and aim to answer the question whether the ``best approximating'' linear model is sufficient. In order to do so, a model with predictor function
\begin{equation}
\label{Su}
\eta(\bs{x}_i) = \beta_1 x_{i1} +... + \beta_p x_{ip} + tr(\bs{x}_i) \ ,
\end{equation}
where the function $tr(\cdot )$ is determined by a tree structure, is fitted. As the tree structure contains the same set of covariates also included in the linear part of the model, the tree is expected to uncover possible non-linear effects and interactions not captured by a linear predictor. 
To determine the optimal model, the data is split into a training and a test sample. After the Model~\eqref{Su} is fitted on the training sample, the sequence of nested subtrees is constructed from the tree structure~$tr(\cdot)$  based on the  Akaike information criterion (AIC; \citealp{Akaike1974}). The trees from this sequence are then evaluated on the test sample again using the AIC. If the selected tree structure contains at least two nodes, the linear model is shown to be insufficient, otherwise the linear model provides a reasonable fit. Unlike the DENDI algorithm, \citet{Su2008} focus on the overall (multivariable) model, but do not give guidance regarding appropriate functional forms for individual covariates. 

The framework of \textit{fractional polynomials} by \citet{Royston1994} is also of particular interest in terms of selecting suitable functional forms of continuous covariates. A fractional polynomial of degree $d$ for variable $x_j$ is defined by 
\begin{equation}
f(x_j) = \xi_0 +\sum_{i = 1}^{d} \xi_i x_j^{p_i} \, ,
\end{equation}
where $p_i \in \{-2, -1, -0.5, 0, -0.5, 1, 2, 3\}$ and $x_j^{p_i} = \log (x_j)$ if $p_i = 0$ for $i = 1,...,d$. If $d=2$ and $p_1 = p_2$, function $f(\cdot )$ is given by $f(x_j) = \xi_0 + \xi_1x_j^{p_1} + \xi_2 x_j^{p_1}\log (x_j)$. \citet{Royston1994} state that for most applications $d \leq 2$ is sufficient resulting in already 8 ($d = 1$) and 36 ($d = 2$) different modeling alternatives, respectively, for a continuous covariate, which offers more flexibility than conventional polynomials.
The choice of the powers $p_i$ is based on the deviance, and model building consists of three steps: (1) testing the overall association of the covariate with the outcome, (2) examining the evidence for nonlinearity, and (3) choosing between a simpler term with degree $d = 1$ and a more complex term with $d = 2$.  This approach aims to determine a suitable function which fits the data well, while being simple, interpretable and generally usable. While fractional polynomials are a flexible and strong tool for detecting nonlinearity, they only consider the covariates separately and neglect possible interactions.

The \textit{multivariate adaptive regression splines} (MARS; \citealp{Friedman1991}) approach is a tool also designed for flexible modeling in generalized regression. It is based on so-called hinge functions, which are given by $\max(0,\,x_{ij}-c_j)$ and \mbox{$\max(c_j-x_{ij},\,0)$}, with split point $c_j$ where a structural break in the slope occurs. Applying one pair of hinge functions results in a modified linear effect equivalent to the predictor function in Equation~\eqref{mult}. MARS applies a stepwise procedure for model building based on the residual sum of squares, where in each step a new pair of hinge functions is included in the model formula either additively or by multiplying them with an existing term (potentially creating interaction terms). In a backward selection step, less relevant terms are subsequently removed utilizing the generalized cross validation (GCV) criterion given by 
\begin{equation}
\text{GCV}(\lambda) =  \frac{\frac{1}{n} \sum_{i=1}^{n}(y_i - \hat{y}_i)^2}{n(1 - (m + \lambda (m-1)))^2}\, ,
\end{equation}
where $m$ denotes the number of terms in the model and $\lambda$ denotes a penalty parameter that is commonly set to a value of 2 or 3 \citep{Hastie2009}. 
Therefore, MARS facilitates the inclusion of modified linear effects and is able to inherently detect interactions (similar to TSVC models). A proposed extension of MARS even allows for a relaxation of the normality and independence assumptions of the outcome variable \citep{Stoklosa2018}. However, as interaction effects in MARS are multiplicative combinations of the hinge functions they do not offer an intuitive interpretation, unlike the group of models DENDI is based on. In addition, MARS assumes that there are structural breaks in the slope  from the start, but simple (non-varying) linear effects as well as piecewise constant effects are not considered. 

\citet{Gertheiss2011} suggested a likelihood ratio test for the check of linearity in an ordinal covariate. Their test is based on a mixed model formulation with penalized dummy coefficients of the ordinal covariate. Unlike DENDI, \citet{Gertheiss2011} focus on one ordinal covariate, only, and potential metrically scaled covariates are not considered.

\section{Simulation study}

To assess the performance of DENDI, we considered five univariable scenarios and one multivariable simulation scenario. 
The aims of our simulation study were (i) to examine the ability of the algorithm to correctly identify the functional forms and interactions (in the multivariable scenario) of continuous covariates, and (ii) to investigate how detection rates are affected by sample size and noise. In the following, the term \textit{detection rate} will be used to describe the proportion of times the correct effect was selected and needs to be distinguished from \textit{selection rate} which refers to the selection of any effect.

In each scenario we simulated a continuous outcome variable that was related to one or multiple standard-normally distributed covariates. We considered sample sizes of $n\in \{ 200, 500, 800\}$. The  error terms $\varepsilon_i$ were drawn from a zero-mean normal distribution with standard deviation $\sigma\in\{1, 1.5, 2\}$. In total this resulted in $3 \times 3 = 9$ settings for each of the six scenarios. In each of the settings we performed 100 replications.

\begin{figure}[!t]
\centering
\includegraphics[width = 0.8\textwidth]{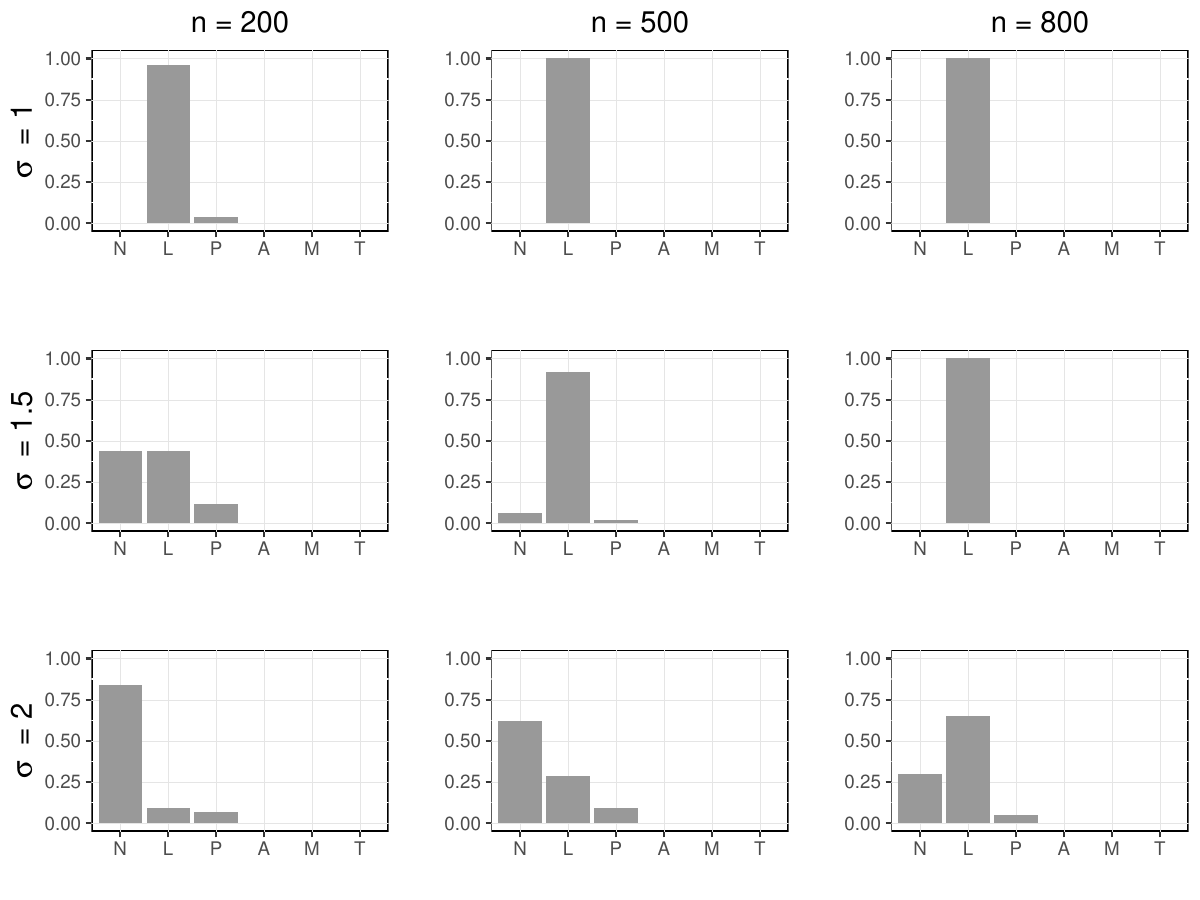}
\caption{Results of the simulation study (scenario 1). The figure shows the proportions of simulation runs in which a null model with no effect~(N), model \eqref{linear} with linear effect~(L), model \eqref{pc} with piecewise constant effect~(P), model \eqref{add} with an additive combination of effects~(A), model \eqref{mult} with a multiplicative combination of effects~(M), and model \eqref{tree} with a tree-structured effect (T) were selected by the algorithm for modeling $x_1$. Selection rates for sample sizes $n \in \{200, 500, 800\}$ (varying across columns) and standard deviations $\sigma \in \{1, 1.5, 2\}$ (varying across rows) are presented.}
\label{fig:SimL}
\end{figure}

\subsection*{Univariable scenarios}

In the following, we consider one covariate $x_1 \sim N(0,1)$. In \textit{scenario 1}, the true data-generating model was the simple linear model \eqref{linear} with $\beta_1=0.5$. The proportions of variance explained by $x_1$ were approximately 0.20 ($\sigma = 1$), 0.10 ($\sigma = 1.5$) and 0.05 ($\sigma = 2$). The results in Figure~\ref{fig:SimL} show that DENDI performed very well in settings with low noise (first row) or high sample size (third column). In three settings ($\sigma=1, n=500$; $\sigma=1, n=800$; $\sigma=1.5, n=800$) the linear effect was correctly identified in all replications. In cases where the linear effect was not found, either no effect or a piecewise constant function~(P) was selected, but none of the more complex alternatives (A, M or T) were identified by the algorithm. In the scenario with high noise and small sample size (\mbox{$\sigma=2, n=200$}) the false negative rate was 0.84, demonstrating a fairly conservative impact of the 1SE rule.   

\begin{figure}[!t]
\centering
\includegraphics[width = 0.8\textwidth]{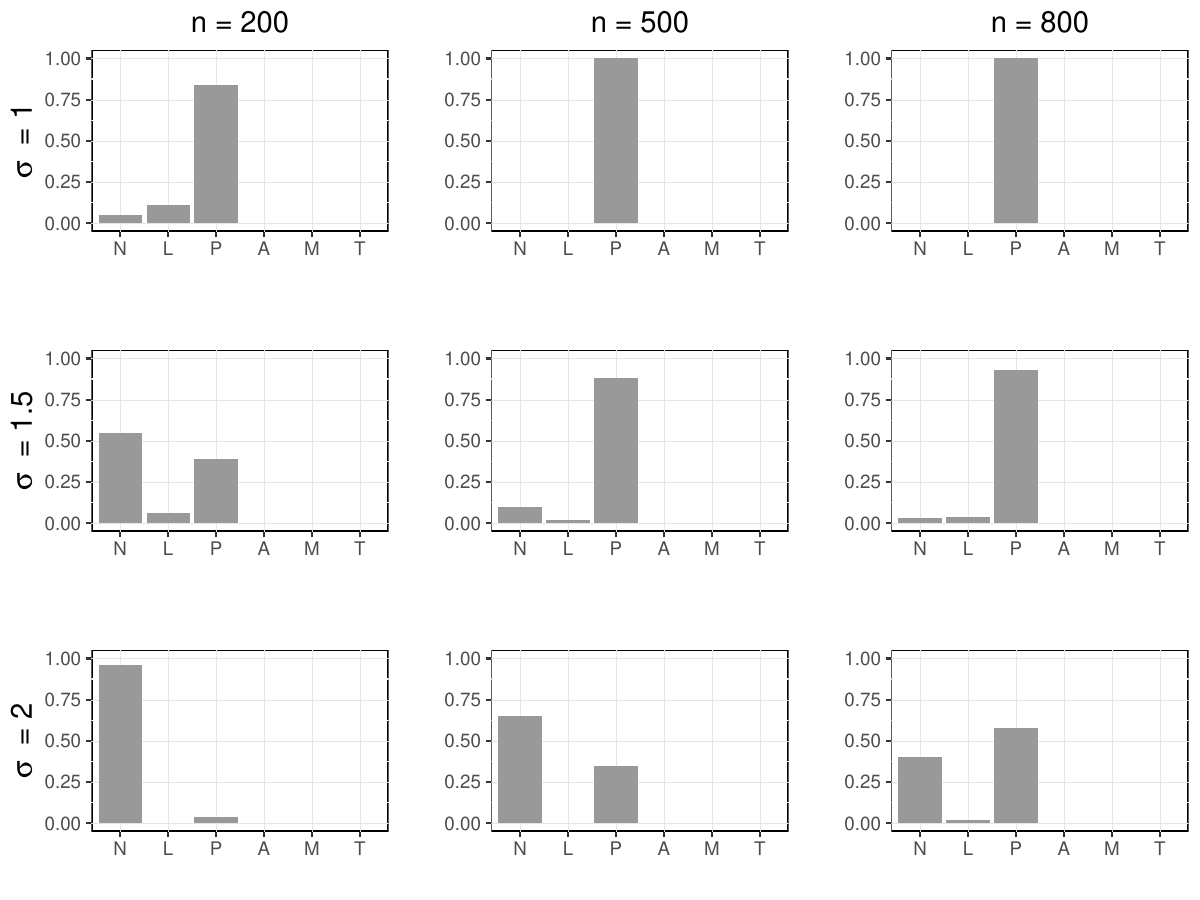}
\caption{Results of the simulation study (scenario 2). The figure shows the proportions of simulation runs in which the different modeling alternatives (N, L, P, A, M and T) were selected by the algorithm for modeling $x_1$. Selection rates for sample sizes $n \in \{200, 500, 800\}$ (columns) and standard deviations $\sigma \in \{1, 1.5, 2\}$ (rows) are presented.}
\label{fig:SimP}
\end{figure}
 
In \textit{scenario 2}, the data was generated according to model \eqref{pc} with predictor 
$\eta(x_{i1}) = \gamma_1 I(x_{i1} > 0)$ with $\gamma_1 = 1$,
which resulted in the same proportions of variance explained by $x_1$ as in scenario~1. Figure~\ref{fig:SimP} shows that the piecewise constant structure was perfectly identified in the scenario with low noise ($\sigma=1$) and medium or high sample size. Summary statistics of the selected split points are given in Table S1 in the Supplement (exemplary shown for $\sigma = 1$ and $n = 800$). In cases where the true effect was not found, again either no effect or a linear effect was selected, but none of the more complex alternatives (A, M or T) were identified by the DENDI algorithm. Compared to scenario 1, the detection rates for the piecewise constant effect were slightly lower than for the linear effect across all settings. This is because two steps have to be performed when fitting the model in \eqref{pc}, namely, finding an optimal split point and estimating the coefficient $\gamma_1$ (and the intercept $\beta_0$), which makes the identification more demanding. 
 
\begin{figure}[!t]
\centering
\includegraphics[width = 0.8\textwidth]{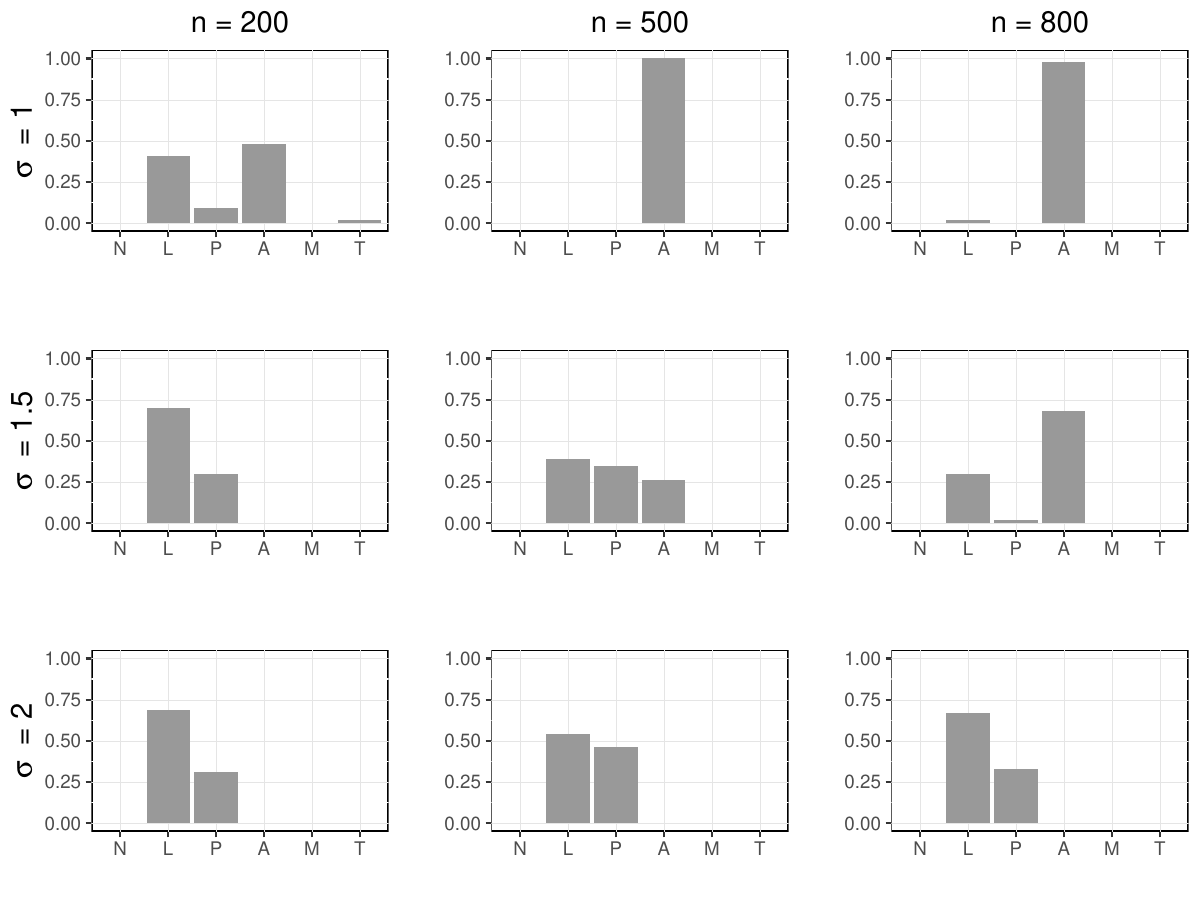}
\caption{Results of the simulation study (scenario 3). The figure shows the proportions of simulation runs in which the different modeling alternatives (N, L, P, A, M and T) were selected by the algorithm for modeling $x_1$. Selection rates for sample sizes $n \in \{200, 500, 800\}$ (columns) and standard deviations $\sigma \in \{1, 1.5, 2\}$ (rows) are presented.}
\label{fig:SimA}
\end{figure}

The data in \textit{scenario 3} was generated according to model \eqref{add} with predictor $\eta(x_{i1}) = \beta_1 x_{i1} + \gamma_1 I(x_{i1}>0)$ with $\beta_1 = 0.7$ and $\gamma_1 = 1.4$. In this scenario and the two following scenarios, the proportions of variance explained by $x_1$ were approximately 0.60 ($\sigma = 1$), 0.45 ($\sigma = 1.5$) and 0.30 ($\sigma = 2$). From Figure~\ref{fig:SimA} it is seen that the true underlying model (A) was predominantly detected in the settings with low noise, only. If noise was large ($\sigma=2$), the additive combination of effects was never found. On the other hand, as the overall effect of $x_1$ was large, some type of effect was always identified (i.e., the false negative rate was zero in all settings). If the additive combination of effects was not identified, either a linear or a piecewise constant function was selected. This may be because for detecting this more complex functional form, the condition of the 1SE rule must be fulfilled twice in the algorithm, requiring strong evidence in favor of these modeling alternatives. 

\begin{figure}[!t]
\centering
\includegraphics[width = 0.8\textwidth]{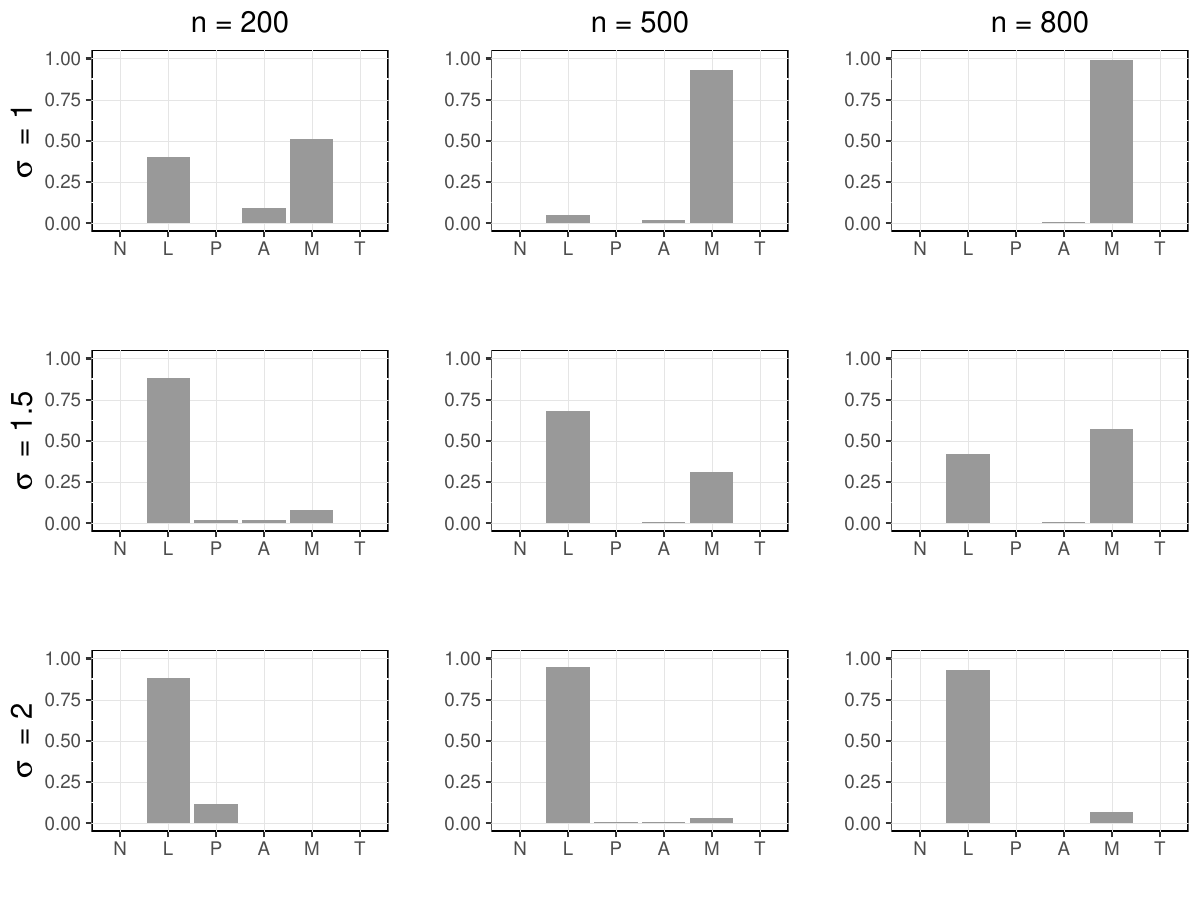}
\caption{Results of the simulation study (scenario 4). The figure shows the proportions of simulation runs in which the different modeling alternatives (N, L, P, A, M and T) were selected by the algorithm for modeling $x_1$. Selection rates for sample sizes $n \in \{200, 500, 800\}$ (columns) and standard deviations $\sigma \in \{1, 1.5, 2\}$ (rows) are presented.}
\label{fig:SimM}
\end{figure}

In \textit{Scenario 4}, the true data-generating model corresponded to \eqref{mult}, which represents a structural break with regard to the slope of $x_1$, and had the form $\eta(x_{i1}) = \beta_{11} x_{i1} + \beta_{12} I(x_{i1}>0)x_{i1}$, with $\beta_{11}=0.6$ and $\beta_{12}=1.2$.
Similar to scenario~3, the true underlying model structure (M) was predominantly detected in the settings with low noise, only (see Figure~\ref{fig:SimM}, upper panel). If the multiplicative combination of effects was not identified, DENDI largely selected a simple linear model \eqref{linear}. For example, in the setting with medium noise and medium sample size ($\sigma=1.5, n=500$) the selection rate for model L was 0.68. The other modeling alternatives (P, A and T) had very low selection rates ($\leq$ 0.12).  

\begin{figure}[!t]
\centering
\includegraphics[width = 0.8\textwidth]{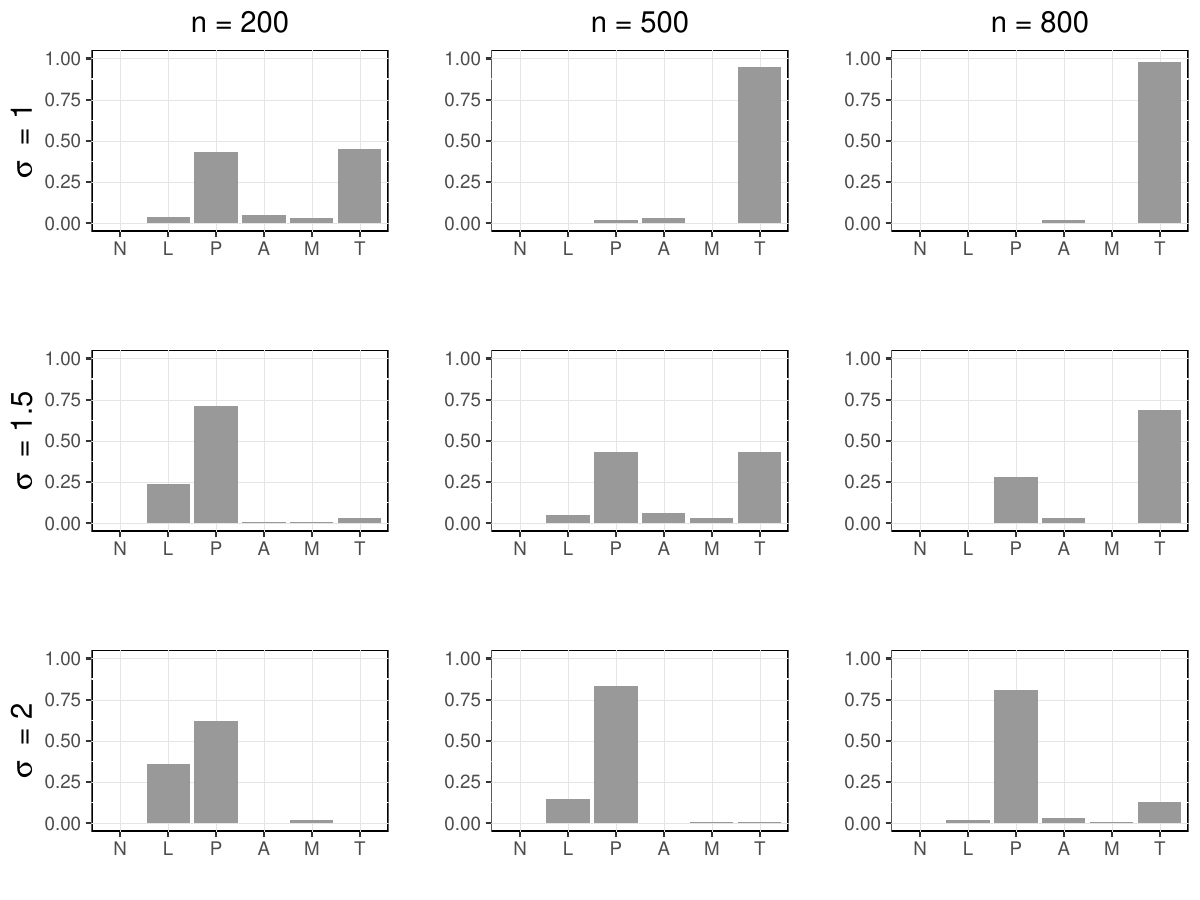}
\caption{Results of the simulation study (scenario 5). The figure shows the proportions of simulation runs in which the different modeling alternatives (N, L, P, A, M and T) were selected by the algorithm for modeling $x_1$. Selection rates for sample sizes $n \in \{200, 500, 800\}$ (columns) and standard deviations $\sigma \in \{1, 1.5, 2\}$ (rows) are presented.}
\label{fig:SimT}
\end{figure}

In \textit{scenario 5}, the data was generated by the tree-structured predictor function $\eta(x_{i1}) = \beta_0 + \gamma_{1\ell} I(x_{i1}\leq 0) + \gamma_{1r} I(x_{i1}>0.675)$ with $\beta_0 = 1$, $\gamma_{1\ell} = -1$, $\gamma_{1r} = 2$. The results shown in Figure~\ref{fig:SimT} are very similar to those of the previous scenario~4. The detection rates for the true underlying effect (T) strongly decrease with increasing noise and decreasing sample size. In the most challenging scenario ($\sigma=2$, $n=200$) model~T was even never selected. Instead, the piecewise constant model (P) was quite often selected, which means that the algorithm performed only one split with regard to~$x_1$. For example, in the setting with high noise and medium sample size ($\sigma=2, n=500$) the selection rate for model~P was 0.83. Note that the linear model (L) was frequently selected, in particular for low sample sizes ($n=200$),  which is likely caused by the monotonic (i.e. non-u-shaped) form of the true underlying effect.

\subsection*{Multivariable scenario}

Lastly, we considered a multivariable scenario with five covariates $x_1,\hdots,x_5 \sim N(0,1)$, where the data was generated by
\begin{equation*}
y_i = 0.6 \cdot x_{i1} + 1.2 \cdot I(x_{i2}>0)x_{i1} + 1 \cdot I(x_{i3}>0) + 2 \cdot I(x_{i3}>0 \land x_{i4}>0 ) + \varepsilon_i\, .
\end{equation*} 
In this scenario, the proportions of variance explained by the covariates were approximately 0.75 ($\sigma = 1$), 0.60 ($\sigma = 1.5$) and 0.45 ($\sigma = 2$). In Table~\ref{tab:Mult} we report the detection rates for the varying effect~\eqref{mult} of $x_1$, the tree-structured effect~\eqref{tree} of $x_3$ and $x_4$ and the null effect of $x_5$. Note that for $x_3$ and $x_4$ the effect was treated as correctly identified if either a tree-structured interaction of $x_3$ with $x_4$ (where the first split was performed in $x_3$) or a tree-structured interaction of $x_4$ with $x_3$ (where the first split was performed in $x_4$) was selected. It is seen that DENDI performed very well across all settings except for the setting with large noise and small sample size ($\sigma=2, n=200$). Overall, the tree-structured interaction between $x_3$ and $x_4$ was more likely to be identified than the varying effect of $x_1$ with regard to $x_2$ (particularly in the settings with large noise). The absence of the effect of $x_5$ was perfectly detected illustrating again the conservative impact of the 1SE rule. More detailed results on the selection rates for the covariates $x_1$, $x_3$ and $x_4$ with regard to each possible effect and interaction are shown in Figures~S1 to S3 in the Supplement. Note that the absence of main effects for $x_2$ and $x_5$ was always identified by the algorithm. 

\begin{table*}[!t]
\caption{Results of the simulation study (multivariable scenario). Proportions of simulation runs in which the effect of $x_1$ (type~\eqref{mult} linear effect modified by $x_2$), the interaction between $x_3$ and $x_4$ (type~\eqref{tree} tree-structured interaction) and the effect of $x_5$ (none) were correctly identified by the algorithm. Detection rates for sample sizes $n\in \{200, 500, 800\}$ and standard deviations $\sigma\in \{1, 1.5, 2\}$ are given. }\label{tab:Mult}
\begin{center}
\begin{footnotesize}
\begin{tabularx}{\textwidth}{X|ccc|ccc|ccc}
\toprule
Effects & \multicolumn{3}{c}{$n = 200$} & \multicolumn{3}{c}{$n = 500$} &  \multicolumn{3}{c}{$n = 800$} \\
 & $\sigma = 1$ & $\sigma = 1.5$ & $\sigma = 2$ & $\sigma = 1$ & $\sigma = 1.5$ & $\sigma = 2$ & $\sigma = 1$ & $\sigma = 1.5$ & $\sigma = 2$  \\ \hline
$x_1$($x_2$) & 0.86 & 0.41 & 0.12 & 1.00 & 0.89 & 0.50 & 1.00 & 0.99 & 0.81 \\
$x_3$, $x_4$ & 0.89 & 0.55 & 0.17 & 1.00 & 0.95 & 0.78 & 1.00 & 0.99 & 0.92 \\
$x_5$ & 1.00 & 1.00 & 1.00 & 1.00 & 1.00 & 1.00 & 1.00 & 1.00 & 1.00 \\
\bottomrule
\end{tabularx}	
\end{footnotesize}
\end{center}		
\end{table*}

Figure~S4 shows the estimated effects when fitting the model using one exemplary data set, where all effects were correctly identified by the proposed algorithm. It is seen that the estimated slopes ($\beta_{11}=0.64$ und $\beta_{12}=1.81$) and the tree-structured effects given in the leaves of the tree in Figure~S4(b) are in line with the true simulated effects. The corresponding TSVC model was fitted using the eponymous R add-on package \textbf{TSVC} \citep{berger2018TSVC}.

\section{Application to the German chronic kidney disease study}

The main objective of the GCKD study was to establish a large cohort of CKD patients who receive comparable medical care and are followed prospectively for up to 10 years. The study enrolled about 5217 patients between 18 and 74 years of age with medium stage CKD. The recruitment period lasted from March 2010 to March 2012. The patients' biomaterials were collected at baseline and at regular intervals during the study \citep{Eckhardt2012, Titze2014}. In our analysis, we included baseline measurements of 3536 patients with an estimated glomerular filtration rate (eGFR) between 30 and 60 mL/min (see below).

\subsection*{Diabetic nephropathy}

Diabetic nephropathy is among the main microvascular complications of diabetes and the leading cause of end-stage kidney disease \citep{Zoja2020}. It refers to the deterioration of kidney function in patients suffering from diabetes mellitus type 1 and type 2. Relevant risk factors for diabetic nephropathy include family history, high blood pressure, dyslipidaemia, obesity, insulin resistance and elevated glycosylated hemoglobin (HbA1c) level \citep{Sulaiman2019}.

To illustrate the DENDI algorithm, we considered the effect of BMI (as an indicator of obesity) and HbA1c level on the probability of suffering from diabetic nephropathy using logistic regression. In the first step of our analysis, we treated the two risk factors separately fitting two univariable models. The results when applying DENDI are shown in Figure~\ref{DN_univ}. The algorithm indicated that for BMI a linear effect is sufficient, while HbA1c exhibits a piecewise constant effect. According to the estimated effects of the corresponding logistic models, the odds of suffering from diabetic nephropathy increases with each BMI point by the factor $\exp(0.09)=1.09$ (see Figure~\ref{DN_univ}(a)), and is $\exp(2.86)=17.46$ times higher for patients with HbA1c level above 49.3 mmol/mol compared to patients with HbA1c level equal to or lower than 49.3 mmol/mol (see Figure~\ref{DN_univ}(b)).

\begin{figure}[!t]
\centering
\begin{subfigure}[b]{0.4\textwidth}
\centering
\includegraphics[width =\textwidth]{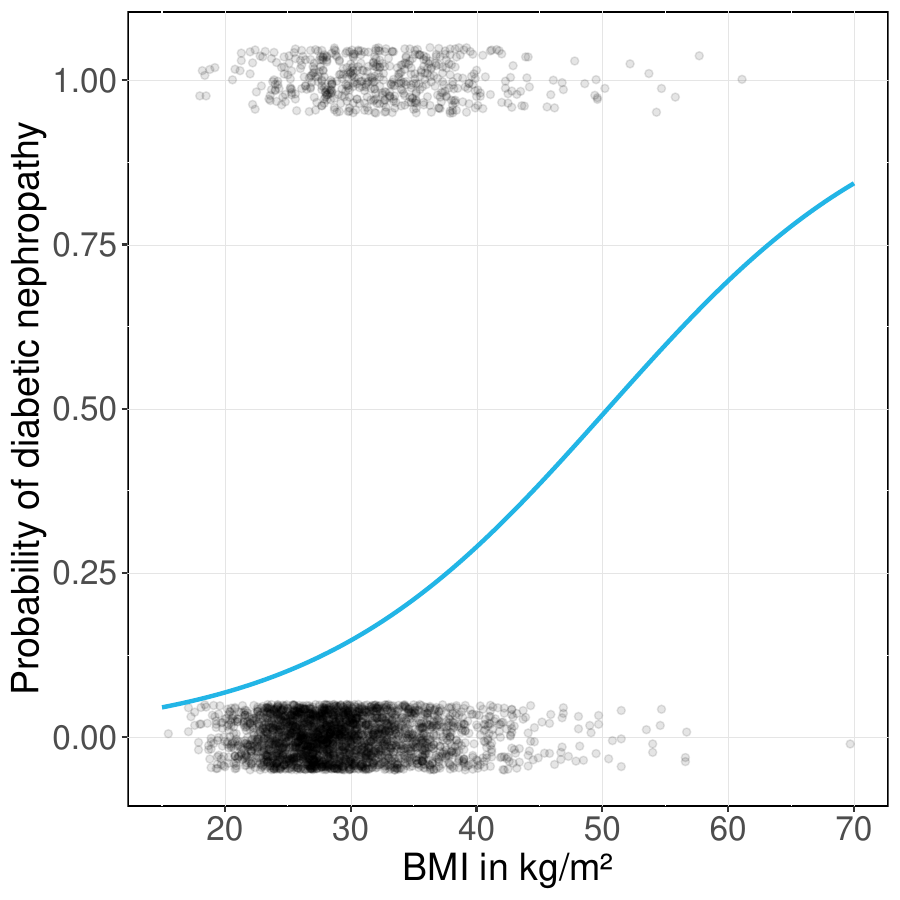}
\caption{Effect of BMI}
\label{BMI_univ}
\end{subfigure}
\begin{subfigure}[b]{0.4\textwidth}
\centering
\includegraphics[width = \textwidth, trim=0cm 0cm 0cm 0cm]{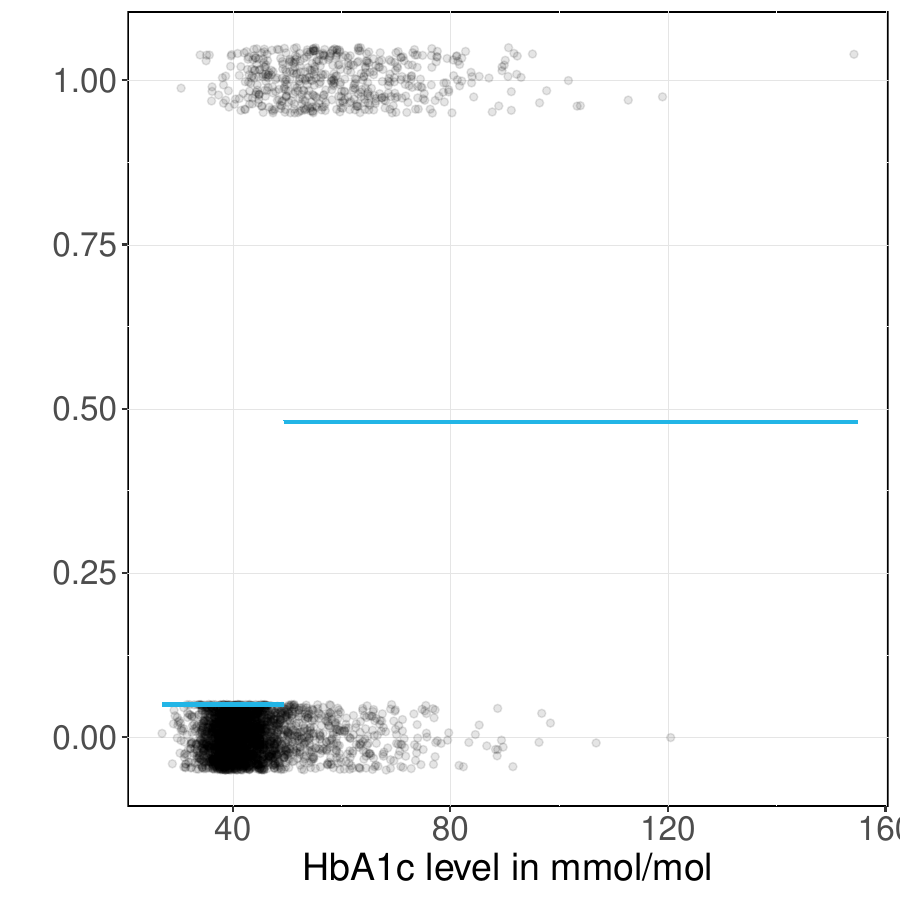}
\caption{Effect of HbA1c level}
\label{HbA_univ}
\end{subfigure}
\hfill
\caption{Analysis of the GCKD study data. Estimated effects of two univariable models for diabetic nephropathy based on the results of DENDI. The left panel~(a) shows the estimated effect of BMI (that is linear on the predictor function) on the conditional probability of suffering from diabetic nephropathy, the right panel~(b) shows the estimated piecewise constant effect of HbA1c. The observed outcome values are marked by jittered black dots.}
\label{DN_univ}
\end{figure}

In a second analysis step, we considered the effects of BMI and HbA1c level in a multivariable analysis additionally adjusting for sex (female or male), educational level (low, intermediate, high or other) and employment status (full-time employed, part-time employed, only domestic work/parenting, retired, seeking work, training/studying or other). The types of the effects for BMI and HbA1c selected by DENDI remained the same as in the separate univariable analyses but the effect sizes changed. The estimated probabilities of suffering from diabetic nephropathy for exemplary patients with mode values of education and employment status as well as median values for BMI and HbA1c level, respectively, are shown in Figure~S5 in the Supplement. The adjusted odds ratio for BMI decreases to $\exp (0.03) = 1.03$ for an increase in BMI by 1 kg/m$^2$, for HbA1c the adjusted odds ratio (with regard to the split point 49.3 mmol/mol) is given by $\exp (2.66)=14.30$. The choice of the piecewise constant effect with one split point is highly clinically meaningful as HbA1c level is commonly used for the diagnosis of diabetes with cut point 48~mmol/ mol~\citep{Nathan2009}.

\subsection*{Estimated glomerular filtration rate}

The GFR is a measure for the severity of CKD recommended by many professional guidelines~\citep{MulaAbed2012}. As radiolabelled methods for measuring GFR are impractical in this study, eGFR (as considered in this paper) was calculated using the  Chronic Kidney Disease Epidemiology Collaboration (CKD-EPI) equation, which is based on the serum creatinine value and accounts for a patient's age, sex and ethnicity \citep{Inker2012}. In this part of the analysis we investigated the association between eGFR and serum hemoglobin levels and urea values, which were identified as influential factors previously~\citep{LopezGiacoman2015, Cao2022}. 

As for diabetic nephropathy, we first treated hemoglobin and urea separately applying the DENDI algorithm for univariable Gaussian models. The algorithm identified a linear effect for hemoglobin. Fitting a linear model yielded an estimated slope of $\hat{\beta}_{\text{Hb}} = 0.71$ (see Figure~\ref{eGFR_univ}(a)). For urea, a more complex effect (i.e. an effect from the third level in the tree structure in Figure~\ref{Fig1}) was selected.  Specifically, DENDI detected an effect of type~\eqref{mult} indicating that a non-linear continuous function may be suitable to capture the relationship between eGFR and urea. To this end we applied TSVC, fractional polynominals, P-splines, and MARS to fit the non-linear effect.  
The TSVC model was specified as recommended by DENDI. For fractional ploynominals the initial degree was set to $d=2$ and testing was performed at $\alpha$-level $0.05$ with the R add-on package \textbf{mfp} \citep{Heinze2023}. The P-spline was fitted based on ten cubic B-spline functions with a second order difference penalty (with the optimal penalty term determined by GCV) using the R add-on package \textbf{mgcv} \citep{Wood2017}. For the MARS approach, we used the implementation from the R add-on package \textbf{earth} \citep{Milborrow2023} and allowed for products of hinge functions up to degree $2$ and set the penalty parameter for the GCV criterion to $\lambda = 3$.

\begin{figure}[!t]
\centering
\begin{subfigure}[b]{0.4\textwidth}
\centering
\includegraphics[width =\textwidth]{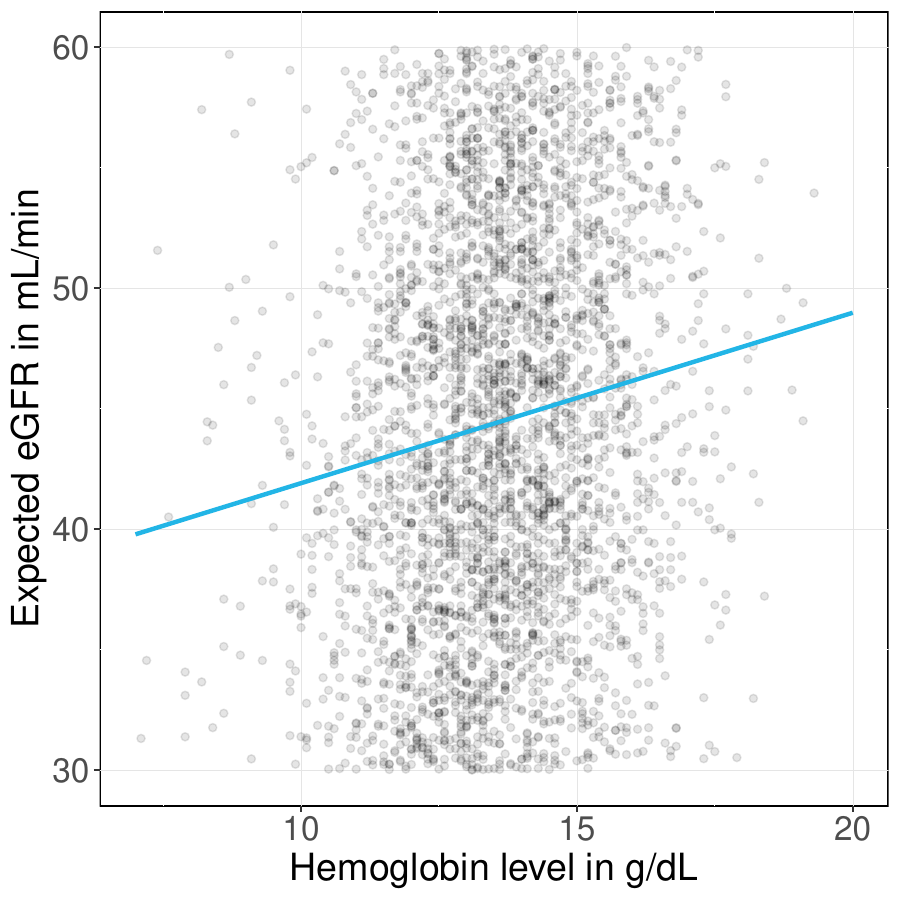}
\caption{Effect of hemoglobin level}
\label{Hemo_univ}
\end{subfigure}
\begin{subfigure}[b]{0.4\textwidth}
\centering
\includegraphics[width = \textwidth, trim=0cm 0cm 0cm 0cm]{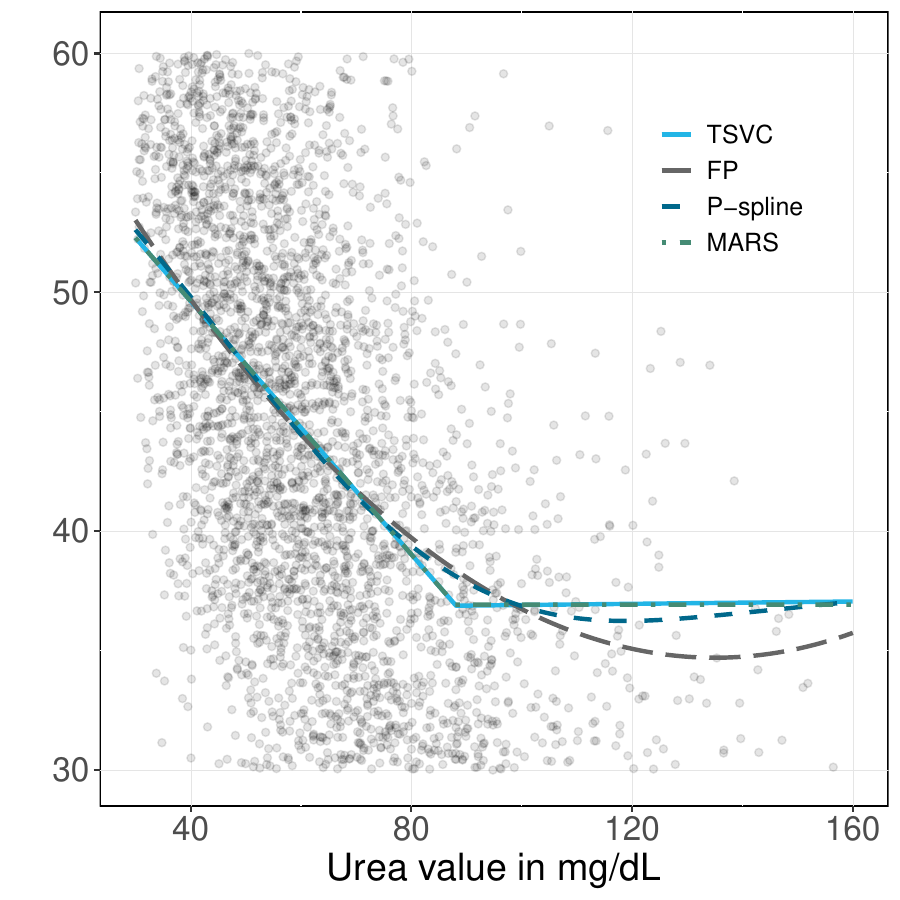}
\caption{Effect of urea}
\label{Urea_univ}
\end{subfigure}
\hfill
\caption{Analysis of the GCKD study data. Estimated effects of univariable analysis for eGFR based on the results of DENDI. The left panel~(a) shows the estimated linear effect of hemoglobin level on the expected eGFR, the right panel~(b) shows more complex modeling alternatives (TSVC, fractional polynomials (FP), P-Spline and MARS) for the effect of urea, where a varying linear effect was detected. The observed outcome values are marked by black dots.}
\label{eGFR_univ}
\end{figure}

Figure~\ref{eGFR_univ}(b) shows that the different approaches yielded very similar fits. According to TSVC a change in the linear effect occurs at an urea value of 88.1 mg/dL, where for urea  $\leq 88.1$ mg/dL the slope was estimated as $\hat{\beta}_{\text{urea,1}} = -0.265$ per 1~mg/dL increase in urea. For urea values larger than 88.1 mg/dL the effect vanishes to $\hat{\beta}_{\text{urea,2}}^* = 0.002$. The MARS approach resulted in a predictor function that comprises only one hinge function with split point $c=87.9$. The fitted fractional polynomial consists of a linear and quadratic term with coefficients $-0.449$ and $0.002$, respectively.

To check whether adjusting for potential confounders changes the identified effects, we considered the effects in a multivariable Gaussian model including educational level and employment status. In analogy to the previous section, Figure~S6 in the Supplement shows the estimated expected eGFR for exemplary patients with mode value of education and employment status as well as median hemoglobin level and urea, respectively. In the multivariable analysis again a linear function was deemed sufficient for the effect of hemoglobin level on eGFR but with a considerably smaller slope ($\hat{\beta}_{Hb} = 0.237$). For urea, the algorithm also selected a nonlinear effect of type~\eqref{mult} again. Therefore, as in the univariable analysis, the shape of the nonlinear effect of urea was considered further. Figure~S6(b) depicts the estimated non-linear functions estimated by the four approaches, which strongly coincide with the univariable analysis. 
 
Overall, our analysis indicates a positive linear effect of hemoglobin level and demonstrates that a simple linear effect is not sufficient to describe the effect of urea on eGFR. These findings are also in accordance to conclusions drawn in previous works \citep{LopezGiacoman2015, Cao2022}.

\section{Summary and discussion}

In this article we propose a detection algorithm that examines various alternatives for modeling continuous covariates and is able to detect different forms of nonlinearity and interactions between covariates, if they are present. The DENDI algorithm is designed to be applied before final model fitting as a tool to facilitate the choice of the general model structure (e.g. a GLM, a GAM or a tree-based approach) and (if a generalized regression model is chosen) to suggest easily interpretable functional forms for individual covariates (e.g.\@ linear functions, structural breaks, categorization as represented by TSVC). The analysis of the GCKD study data demonstrates how the proposed algorithm can be applied to investigate whether linear effects are sufficient or more complex nonlinear effects would be recommended. Our results importantly indicate (i) that dichotomization is highly useful to describe the effect of HbA1c level on the probability of suffering from diabetic nephropathy, and (ii) that the effect of urea value on the expected eGFR is not simply linear. 
The results of the simulation study show that DENDI performs well in univariable analyses as well as in a multivariable scenario. Due to the (repeated) application of the 1SE rule, false positive results are avoided, which means that the complexity of the effects tends to be underestimated (particularly in settings with small sample size and/or large noise).  

The modeling alternatives taken into account by the algorithm are nested within one another and contain (combinations of) linear and piecewise constant effects, as well as two-factor interactions. All the models can be fitted within the framework of TSVC models. Our implementation, which is part of the supplement to this article, makes use of the \textbf{TSVC} package in~R.
Goodness-of-fit tests for parametric regression models were proposed by \citet{Fan2001}, \citet{Shah2018} and \citet{Jankova2020}. The principle is to construct a test statistic based on the residuals to identify how well a model fits the data. These tests are very flexible tools, but the focus is on an overall check of model misspecification rather than on the effects of individual covariates. The closest relation to the DENDI algorithm is to the approach by \citet{Royston1994} as it also investigates the effect of a continuous covariate at a time and chooses between functions of varying complexity. The proposed algorithm differs, however, in the sense that it allows for interactions between the covariates and is based on tree-structured varying effects which are easily accessible and interpretable.
 
In order to avoid sampling issues induced by random splitting of the data, DENDI applies LOOCV to compare the predictive performance of the considered models. This leads to a comutational cost of $\mathcal{O}(np^2)$ as in each of the $n$ LOOCV iterations, $\mathcal{O}(p^2)$ models are fitted. LOOCV is less biased than $k$-fold cross-validation and particularly advantageous for low sample sizes as nearly the whole sample is used for training in each iteration \citep{Elisseeff2005}. Yet, LOOCV is computationally intensive and leads to increased variance, in particular compared to repeated $k$-fold cross-validation \citep{Boulesteix2008}. The comparison of alternative resampling schemes (e.g. repeated cross-validation or bootstrap) may be an interesting topic for further research.

 \subsection*{Conflict of interests}
Declarations of interest:
The authors report there are no competing interests to declare.

\subsection*{Acknowledgements}
Support by the German Research Foundation (DFG), grant BE 7543/1-1, is gratefully acknowledged.

\bigskip
\begin{center}
{\large\bf SUPPLEMENTARY MATERIAL}
\end{center}

\begin{description}

\item[Additional information:] Detailed description of the DENDI algorithm as well as additional tables and figures that illustrate the results of the simulation study and the application. (pdf file)

\item[R-code for the algorithm:] R-code of the functions that perform the algorithm. (R file)

\end{description}

\bibliographystyle{Chicago}
\begin{spacing}{1.4}
\bibliography{Manuscript}
\end{spacing}
\end{document}